\documentclass[aps,prc,twocolumn,tightenlines,superscriptaddress,nofootinbib,a4paper,showpacs]{revtex4}
\usepackage{dcolumn}
\usepackage{graphicx}
\usepackage{rotating}
\usepackage{natbib}
\usepackage{amsmath}
\usepackage{braket}
\begin{document}

\title{Systematics of intermediate-energy single-nucleon removal cross sections}

\author{J.~A.~Tostevin}
\affiliation{Department of Physics, Faculty of Engineering and
Physical Sciences, University of Surrey, Guildford, Surrey GU2 7XH,
United Kingdom}
\affiliation{National Superconducting Cyclotron Laboratory, Michigan
State University, East Lansing, Michigan 48824, USA}

\author{A.~Gade}
\affiliation{National Superconducting Cyclotron Laboratory, Michigan
State University, East Lansing, Michigan 48824, USA}
\affiliation{Department of Physics and Astronomy, Michigan State
University, East Lansing, Michigan 48824, USA}
\date{\today}

\begin{abstract}
There is now a large and increasing body of experimental data and theoretical
analyses for reactions that remove a single nucleon from an intermediate-energy
beam of neutron- or proton-rich nuclei. In each such measurement, one obtains
the inclusive cross section for the population of all bound final states of the
mass $A-1$ reaction residue. These data, from different regions of the nuclear
chart, and that involve weakly- and strongly-bound nucleons, are compared with
theoretical expectations. These calculations include an approximate treatment
of the reaction dynamics and shell-model descriptions of the projectile initial
state, the bound final states of the residues, and the single-particle strengths
computed from their overlap functions. The results are discussed in the light
of recent data, more exclusive tests of the eikonal dynamical description, and
calculations that take input from more microscopic nuclear structure models.
\end{abstract}

\pacs {24.50.+g, 24.10.Ht, 25.60.-t, 25.70.-z}
\maketitle

Fast nucleon removal reactions have been developed as an effective direct
reaction, producing highly neutron-proton asymmetric nuclei with relatively
high cross sections. The combination of intermediate-energy secondary beams
and thick reaction targets has led to precise measurements for a large number
of the most exotic nuclei. The cross sections are usefully large because: (a)
essentially all bound reaction residues are detected, with $\approx$100\%
efficiency, and (b) the measured cross sections are highly inclusive -- with
respect to the target final states. We concentrate here on the systematics
of such reactions on light target nuclei, either $^9$Be or $^{12}$C, for
which nucleon removal associated with the Coulomb interaction (i.e. elastic
Coulomb dissociation) is negligible. Here, the two strong-interaction-driven
nucleon-removal mechanisms are elastic and inelastic breakup of the projectile
in which the target nucleus remains in or is excited from its ground state,
respectively \cite{Han03}.

In this work we discuss, in the main, the sum of these two contributions. In
measurements that determine only the number of bound residual nuclei, the cross
sections are, of course, also inclusive with respect to all bound states of
the reaction residue. In such proton-neutron asymmetric systems, these final
state spectra are very often unknown or only partially known. In the following
analyses these final state spins, parities and excitation energies are therefore
normally taken from shell-model calculations and the theoretical inclusive cross
sections are taken to be the sum of the calculated cross sections to all of the
predicted shell-model states of the residue with excitation energies below the
empirical, if known, or the evaluated \cite{Aud03} first nucleon separation energy.

The eikonal model theoretical description of the nucleon removal reaction dynamics,
that uses the sudden (fast collision) and eikonal (forward scattering) approximations,
has been presented and discussed in detail elsewhere, e.g. Refs. \cite{Tos01,Han03}
and references therein. The model assumes that for the fast, surface-grazing
interactions of the mass $A$ projectile with the target, of relevance to the single
nucleon removal channel, the state $\alpha$ of the mass $A-1$ reaction residue is
a spectator. Thus, the yield of residues in a particular final state $\alpha$
reflects a component (a parentage) of this configuration in the ground-state
wave function of the projectile.

The partial cross section for removal of a nucleon, from a single-particle
configuration $j^{\pi}$, populating the residue final state $\alpha$ with excitation
energy $E_\alpha^*$, is calculated as
\begin{equation}
\sigma_\text{th}(\alpha)=\left(\frac{A}{A-1}\right)^{\!\!N} C^2S(\alpha,j^{\pi})
\,\sigma_\text{sp}(j,S_\alpha^*) \label{eq:xsec} ,
\end{equation}
where $S_\alpha^*=S_{n,p}+E_\alpha^*$ is the effective separation energy
for the final state $\alpha$ and $S_{n,p}$ is the ground-state to ground-state
nucleon separation energy. Here $N$, in the $A$-dependent center-of-mass
correction factor that multiplies the shell-model spectroscopic factors
$C^2S(\alpha,j^{\pi})$, is the number of oscillator quanta associated with
the major shell of the removed particle \cite{Die74}. The single-particle
cross section $\sigma_\text{sp}$ is the sum of the elastic and inelastic
breakup contributions to the reaction \cite{Tos01}, $\sigma_\text{sp}=
\sigma_\text{sp}^\text{inel} + \sigma_\text{sp}^\text{elas}$, calculated
assuming the removed-nucleon's single-particle wave function (or overlap) is
normalized.

Thus, the theoretical inclusive nucleon-removal cross section, $\sigma_\text{th}$,
is calculated as the sum of these partial cross sections $\sigma_\text{th} (\alpha)$
for all bound shell-model final states of the mass $A-1$ residue. This theoretical
cross section is therefore an overall measure of the predicted reaction yield resulting
from the single-particle strengths of the low-energy, shell-model spectrum of the
nucleus. There are numerous inputs to the $\sigma_\text{th}$ calculation (discussed
below), that specify the ranges of the optical potentials and the nucleon radial
overlaps, and hence dictate the reaction geometry, and the inclusive cross section
does not directly or simply probe the value of any individual spectroscopic factor,
$C^2S(\alpha,j^{\pi})$. The overall shell-model strengths and reaction yields can
however be compared with the measured cross sections, $\sigma_\text{exp}$. This
comparison is usually made in terms of the cross sections ratio $R_s = \sigma_
\text{exp}/\sigma_\text{th}$.

As a measure of the asymmetry of the neutron and proton binding, and that of their
Fermi surfaces (that strongly affects the absolute cross sections), we use the
parameter $\Delta S$. If there is just one populated final state $\alpha$, the
residue ground state, then $\Delta S = S_n-S_p$ for neutron removal and $\Delta
S =S_p-S_n$ for proton removal. When there are several residue final states populated
then the separation energy of the removed particle in $\Delta S$ is replaced by
the weighted average of their $S_\alpha^*$, each weighted by their calculated
partial cross sections, $\sigma_\text{th}(\alpha)$. With this convention, the removal of
the most strongly-bound (weakly-bound) nucleons from proton-neutron asymmetric
nuclei have large positive (negative) values of $\Delta S$.

For each projectile, the calculation of $\sigma_\text{th}$ involves several
inputs: (i) realistic shell-model calculations, spectra and $C^2 S$ values,
(ii) realistic residue- and nucleon-target complex optical potentials and their
derived elastic scattering S-matrices, that enter the eikonal model impact parameter
integrals for $\sigma_\text{sp}^\text{inel}$ and $\sigma_\text{sp}^\text{elas}$
\cite{Tos01} and localize the reactions spatially, and (iii) realistic geometries
for the radial wave functions (overlap functions) for the initial bound states
of the removed nucleons in the projectile ground state. In exotic nuclei, many
of these parameters are uncertain. The strategy used in the analyses discussed
here is to employ the best available shell-model calculations for (i), and to
constrain the shapes and radial size parameters of the optical potentials, in (ii),
and overlaps, in (iii), by the use of theoretical model systematics; specifically,
by the use of Hartree-Fock (HF) calculations of neutron and proton densities
for the residues and the root mean squared (rms) radii of orbitals in the HF
mean-field. The procedure used, applied to all of the data sets shown here, is
detailed in Ref. \cite{Gad08}. We note that, for most of the data sets, which
are for beam energies near 100 MeV/nucleon on a $^9$Be target, the neutron-
and proton-target potentials and their (eikonal) elastic S-matrices are in
fact essentially common to the analyses of a large number of data sets for
reactions for projectiles with a wide range of nucleon separation energies.

\begin{figure}[t]
\begin{center}
\includegraphics[width=1.0\columnwidth,angle=0,scale=1]{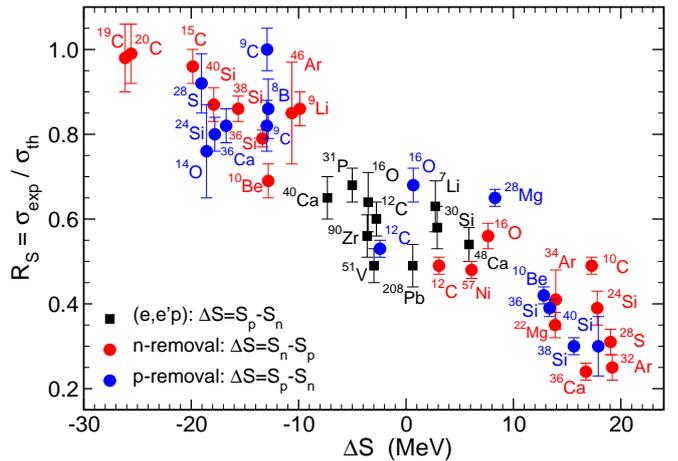}
\end{center}
\caption{(Color online). Compilation of the computed ratios $R_s$ of the
experimental and theoretical inclusive one-nucleon removal cross sections
for each of the projectile nuclei indicated. $R_s$ is shown as a function of
the parameter $\Delta S$, used as a measure of the asymmetry of the neutron
and proton Fermi-surfaces. The red points are for neutron removal cases and
the blue points those for proton removal. The solid (black) squares, deduced
from electron-induced proton knockout data, are identical to the earlier
compilation of Ref. \cite{Gad08}.}
\label{fig:one}
\end{figure}

The first consistent analyses using this approach for data involving
the removal of a well-bound nucleon, e.g. a neutron with separation energy
$S_n\approx 22$ MeV from the proton-rich nucleus $^{32}$Ar \cite{Gad04}, now
denoted $^{32}$Ar($-n$) with $\Delta S \approx +20$ MeV, showed that the cross
section ratio $R_s$ was unexpectedly small, with $R_s$=0.24(3). Reactions involving
weakly-bound nucleons, on the other hand, e.g. the $^{15}$C($-n$) reaction with
$S_n$=1.22 MeV and $\Delta S \approx -20$ MeV, were consistent with $R_s$ values
near unity \cite{Ter04}. A first systematic analysis and compilation of
available data was presented in 2008, in Fig. 6 of Ref. \cite{Gad08}. This
incorporated a previous analysis \cite{Bro02} of existing high-energy data
for the $^{12}$C($-n,-p$) and $^{16}$O($-n,-p$) reactions, that showed
consistency, for these stable nuclei, with analogous $R_s$ values deduced
from high-energy electron-induced proton knockout. These data points, with
relatively small $|\Delta S|$, are clustered near the center of Fig.\
\ref{fig:one}. These suppressed $R_s$ values, from many electron-induced
proton knockout studies on stable nuclei, have been carefully studied and
quantified, see e.g. the review of Ref. \cite{Dic04}. Principally, these
result from nucleon single-particle strengths in low-lying shell-model
configurations being depleted due to their mixing: (a) with higher-lying
shells, by correlations involving the strong short-range behavior of the
nucleon-nucleon interaction, and (b) with more collective configurations
involving surface and/or volume correlations of longer range. Exotic beam
data have allowed an exploration of the behavior of $R_s$ for a much-extended
range of $|\Delta S|$ values away from the stable nuclei, and to include
both neutron and proton removal reactions.

A compilation of the results of the (residue bound-states-inclusive) data
and analyses, that use the common eikonal model calculations with shell-model
effective interactions and model spaces appropriate to the $(N,Z)$ of the
system, are shown as calculated $R_s=\sigma_\text{exp}/\sigma_\text{th}$
values versus $\Delta S$ in Fig.\ \ref{fig:one}. Here, the reactions data
shown in the earlier Fig. 6 of Ref. \cite{Gad08} are supplemented
by the measurements and analyses for $^{57}$Ni($-n$) \cite{Yur06}, $^{22}
$Mg($-n$) \cite{Dig08}, $^{9}$Li($-n$), $^{9}$C($-p$), $^{10}$Be($-n,-p$),
$^{10}$C($-n$) \cite{Gri12}, $^{36}$Ca($-n,-p$) \cite{Sha12}, $^{19,20}
$C($-n$) \cite{Kob12}, $^{36,38,40}$Si($-n,-p$) \cite{Str14}, $^{28}$Mg($
-p$) \cite{Wim14}, and $^{14}$O($-p$) \cite{Sun14}. The value of $R_s$ for
this latter $^{14}$O($-p$) data point (measured on a carbon target), with its
relatively large error bar, has been recalculated here to be consistent
with the HF methodology used for the other analyses. This single-particle
cross section is calculated to be 27.76 mb. So, based on the reported
$\sigma_\text{exp} =35(5)$ mb, when using the ground-state to ground-state
spectroscopic factor $C^2S=1.55$ of the WBT shell-model interaction (e.g.
Table I of \cite{Fla12}), we deduce that $R_s$ = 0.76(11), as shown in
Fig.\ \ref{fig:one}. The value is smaller than, but is consistent with,
the value estimated in Ref. \cite{Sun14}.

It should also be noted that
the $^{10}$Be, $^{36}$Ca and $^{36,38,40}$Si cases, as for the earlier
$^{28}$S($-n,-p$) and $^{24}$Si($-n,-p$) data of Ref.\ \cite{Gad08}, include
data for the removal of nucleons of both the excess (weakly-bound) and
the depleted (strongly-bound) species from the same projectile, with
experimental (systematic uncertainty) advantages. Compared to
the earlier compilation, Fig.\ \ref{fig:one} now includes several reactions
with large positive $\Delta S$ that involve proton removal. While there is
some degree of scatter on the individual points for the different projectile
masses, which use shell-model analyses made with different effective
interactions and/or model spaces, the trend of this large body of data is
remarkably consistent. We note that the largest scatter and departures from
the (nominally linear) trend, for the larger $|\Delta S|$, tend to involve
the lighter nuclei studied where individual departures from the HF- and
shell-model-based calculations used are not unexpected.

The large majority of points in Fig.\ \ref{fig:one} are based on measurements
made at the National Superconducting Cyclotron Laboratory (NSCL), at Michigan State
University, with secondary beam energies in the range 80-100 MeV/nucleon incident
on a $^9$Be target. Exceptions are the $^{19,20}$C($-n$) data (extreme left),
measured at the RIBF, RIKEN at 240 MeV/nucleon \cite{Kob12}, the $^{14}$O($-p$)
data point (with $\Delta S=-18.57$ MeV), measured at the HIRFL, Lanzhou at
305 MeV/nucleon \cite{Sun14}, and the precision stable beam data from Berkeley,
for the $^{12}$C($-n,-p$) and $^{16}$O($-n,-p$) reactions at 250, 1050 and 2100
MeV/nucleon (center); as were analyzed in Ref.\ \cite{Bro02}. All of these higher
energy data were measured on a carbon target. The $^{10}$C($-n$) and $^{10}$Be($
-n$) reaction points (far-right and left of center, respectively) were measured
at the NSCL at 120 MeV/nucleon \cite{Gri12}.

The intermediate energy of the beam is important, generally, for the applicability
of the sudden and eikonal dynamical approximations used. The beam energy is also
of particular importance for the points with large positive $\Delta S$ on the far
right hand side of the Figure. These involve the removal of the most strongly-bound
nucleons. Four-momentum conservation in an endoergic nucleon-removal reaction from
the projectile will naturally impose a kinematic upper-limit on the longitudinal
momentum carried by the fast, forward-travelling reaction residues. As was shown
dramatically in Ref. \cite{Fla12}, where such a well-bound case, the $^{14}$O($-n$)
reaction with $S_n= 23.2$ MeV, was performed with too low a secondary beam energy
(in that case only 53 MeV/nucleon), this kinematical cutoff intruded into and
severely distorted the momentum distribution of the $^{13}$O residue cross section.
An expression for this maximal beam-direction residue momentum, $P_\|$, said to
be consistent with energy and momentum conservation, was given in Eq. (1) of
\cite{Fla12}. This was
\begin{equation}
P_\|=\left[(T_p-S_n-\epsilon_f)^2+2M_r(T_p-S_n-\epsilon_f)
\right]^{1/2}\ , \label{eq:one}
\end{equation}
where $\epsilon_f$ is the neutron-target relative energy in the final state.
However, this expression, that makes no reference to the target mass, $M_t$,
originates from a model that assumes $M_t$ is infinitely massive, and should
not be used for kinematics calculations with a light (e.g. $^9$Be) target.
The correct kinematics is more complex. With $E_i = M_i+T_i$ and $P_i$ the
laboratory-frame energies and momenta, and $M_i$ the rest energies of the
projectile ($p$), target ($t$), reaction residue ($r$) and removed nucleon
($N$), and assuming a final-state excitation energy $\varepsilon_f^*$  in
the removed-nucleon+target system, four-momentum conservation actually
requires that
\begin{eqnarray*}
M_p^2+M_t^2+M_r^2+2M_tE_p-2\left( [E_p+M_t] E_r-P_p P_\|\right)\nonumber
\\\ \ \ = \left(M_t+M_N+\varepsilon_f^*\right)^2 \ .\ \ \ \ \ \ \ \
\end{eqnarray*}
The general conclusion of Ref. \cite{Fla12} is however confirmed; that this
kinematic cutoff has minimal effect for reactions with beam energies near to
and in excess of 80 MeV/nucleon. It is also clear from \cite{Fla12}, that
any distortion observed on the high-momentum side of the (usually Gaussian-like)
residue momentum distribution provides a valuable diagnostic for the onset
of such effects.

For these reasons, it would certainly be of interest and value for one or more
of the cases involving removal of the most well-bound nucleons to be measured
at higher beam energies, say in excess of 200 MeV/nucleon. For example, the
cross section for the $^{10}$C($-n$) reaction at 1.6 GeV/nucleon, with $S_n=
21.28$ MeV, has been reported informally to be 21.4(17) mb \cite{Vol11}. The
residue in this case, $^{9}$C, has only one bound state and at 1.6 GeV/nucleon
the calculated $\sigma_\text{sp}$ is 22.8 mb. When using the Cohen and Kurath
shell-model effective interaction \cite{CoK65}, as used and tabulated in Table
V of \cite{Gri12}, the derived $R_s$ is 0.49(4), in excellent agreement with
the value 0.49(2) from the 120 MeV/nucleon reaction analysis of Ref. \cite{Gri12}.
Thus, this 1.6 GeV/nucleon measurement coincides with and confirms the
$^{10}$C($-n$) data point near the right hand edge of Fig.\ \ref{fig:one}.
Further checks of this sort, for the $sd$-shell and heavier nuclei in this
region of $\Delta S$ in the figure, would clearly be of value.

There have been several studies to try to understand the behavior of $R_s$
with $\Delta S$ and to test the theoretical inputs to the $\sigma_\text{th}$.
To test the approximate description of the reaction mechanism, two recent studies
have probed the relative importance of the two contributions to the removal
cross section as calculated using the eikonal model, that is $\sigma_\text{sp}
^\text{inel}$ and $\sigma_\text{sp}^\text{elas}$. These used more exclusive
measurements of the final state, specifically, coincidences of the reaction
residues with light charged particles. These data and analyses, that looked
at both weakly-bound, $^8$B($-p$) and $^9$C($-p$) \cite{Baz09}, and well-bound,
$^{28}$Mg($-p$) \cite{Wim14}, proton removal cases, are in good agreement with
the relative magnitudes of the cross sections calculated with the eikonal model.

Another approach has been to try to interface the reaction dynamics with
structure models that go beyond the truncated-basis shell-model - that is
to take the residue densities, nucleon radial overlaps and their spectroscopic
factors from more microscopic calculations. In Ref. \cite{Gri11}, this procedure
was adopted for the light projectile reactions $^{10}$Be($-n$) and $^{10}$C($-n$).
There, based on the no basis, variational Monte Carlo (VMC) \cite{Pie01} structure
information, the shell-model $R_s$ values of Fig. \ref{fig:one}, of 0.69(4) and
0.49(2), respectively, became $R_s^\text{VMC}$ of 1.00(5) and 0.75(4), with
significant changes from the restricted $p$-shell shell-model calculations.
Since, in addition to the more extended basis, the VMC calculations include
both a realistic nucleon-nucleon (NN) interaction and a model three-nucleon
(3N) interaction, the most important of these physical ingredients could not
be determined. In the neutron-rich oxygen isotopes, coupled-cluster calculations
\cite{Jen11} have also pointed to the potential importance of couplings of the
near-threshold neutrons to their continuum upon the spectroscopic factors of
the overlaps for proton removal from the well-bound Fermi surface. However,
as is shown in Eq. (1), the $C^2S$ are just one ingredient to $\sigma_\text{th}$
and $R_s$ and the spatial extent and the radial forms of these proton overlaps,
and also the residue densities, are required for a consistent calculation of
these effects upon the removal reaction cross sections. All of these ingredients
and comparative cross section calculations should be possible soon.

Such efforts have concentrated on the suppressed $R_s$ values for large positive
$\Delta S$ in the Figure. There has been very little discussion of the larger
$R_s$ with large negative $\Delta S$, where it is more natural to associate
the weak binding and larger radial extent of the nucleon orbitals with reduced
correlation effects beyond the shell-model. This is in keeping with the trends
observed of the coupled-cluster calculations of Ref. \cite{Jen11}.

In this Brief Report we have brought up-to-date a comparison of measured and
calculated inclusive one-nucleon-removal cross sections, in the light of analyses
of a large body of new experimental data and recent more exclusive measurements
that allow additional tests that add confidence to the predictions of the reaction
mechanism. The new data conform to the earlier trends of the ratio of the measured
and theoretical cross sections, $R_s = \sigma_\text{exp}/\sigma_\text{th}$, with
the neutron-proton separation energy asymmetry parameter $\Delta S$. Recent
theoretical results, based on structure models that go beyond the truncated-basis,
configuration-interaction shell-model, suggest that extended bases, 3N force
effects, and explicit couplings of near threshold single-particle configurations
to the continuum may all play a role in understanding the observed dependence
of $R_s$ on $\Delta S$. The present eikonal reaction plus shell-model structure
theory approach predicts that more single-particle strength and nucleon-removal
cross section leads to bound final states of the residues than is observed, in
both neutron and proton removal, particularly in reactions that remove very
well-bound nucleons of the deficient species. The new data for weakly-bound
neutron and proton removals with large negative $\Delta S$ have $R_s$ values
consistent with unity, with little evidence of the need for any significant
correlation effects beyond the shell-model.

This work was supported by the National Science Foundation under Grant No. PHY-1102511
and the United Kingdom Science and Technology Facilities Council (STFC) under Grants
Nos. ST/J000051/1 and ST/L005743/1.

\bibliographystyle{apsrev}

\end{document}